\documentclass[conference]{IEEEtran}
\usepackage{cite}
\usepackage{hyperref}

\usepackage[pdftex]{graphicx}
\DeclareGraphicsExtensions{.pdf}
\usepackage[usenames,dvipsnames,svgnames,table]{xcolor}
\usepackage[cmex10]{amsmath}
\usepackage{amsfonts}
\usepackage{array}
\usepackage{color}
\usepackage{mdwmath}
\usepackage{mdwtab}
\usepackage[caption=false,font=footnotesize]{subfig}
\usepackage{fixltx2e}
\usepackage{url}

\usepackage{booktabs}

\begin{document}
\title{A Coverage Based Decentralised Routing Algorithm for Vehicular Traffic Networks}

\author{\IEEEauthorblockN{Timothy Barker, Chao Zhai, \& Mario di Bernardo}
\IEEEauthorblockA{Department of Engineering Mathematics, \\
University of Bristol,
Bristol, UK\\
Email: timothy.barker@bristol.ac.uk, chao.zhai@bristol.ac.uk, m.dibernardo@bristol.ac.uk}}
%\and
%\IEEEauthorblockN{Chao Zhai}
%\IEEEauthorblockA{Department of\\Engineering Mathematics\\
%University of Bristol\\
%Bristol, UK\\
%Email: chao.zhai@bristol.ac.uk}
%\and
%\IEEEauthorblockN{Mario di Bernardo
%\IEEEauthorblockA{Department of\\Engineering Mathematics\\
%University of Bristol\\
%Bristol, UK\\
%Email: enmdb@bristol.ac.uk}}}

%Telephone: (800) 555--1212, Fax: (888) 555--1212}


\maketitle

\begin{abstract}
We present a simple yet effective routing strategy inspired by coverage control, which delays the onset of congestion on traffic networks, by introducing a control parameter. The routing algorithm allows a trade-off between the congestion level and the distance to the destination. Numerical verification of the strategy is provided on a number of representative examples in SUMO, a well known micro agent simulator used for the analysis of traffic networks. We find that it is crucial in many cases to tune the given control parameters to some optimal value in order to reduce congestion in the most effective way.  The effects of different network structural properties are connected to the level of congestion and the optimal range for setting the control parameters.
\end{abstract}

% INTRODUCTION
\section{Introduction}
The ability of traffic networks to support an increasing amount of vehicular traffic is becoming crucially important, as there are social, environmental and economic consequences of poorly managed networks \cite{Stokols1978, Barth2008, Graham2007}. In particular, vehicular congestion significantly jeopardizes the performance of traffic networks. For instance, there is evidence to suggest that in cities such as London, travel delays during busy periods increase journey times by an average of 34\% \cite{tomtom2014}. Also, a recent analysis of road speeds in New York suggests that it suffers from a rush hour which lasts all day \cite{Wellington2014}. There are a number of factors that may contribute to congestion on a traffic network. The most obvious is an increase in the number of vehicles using the network exceeding its capacity, for example during peak periods \cite{dot2014}. If some roads become congested, congestion will tend to spread to other parts of the network, so the faster those roads can be cleared, the better the network performs. With car ownership in the UK having increased over the past 40 years \cite{dft2014}, this problem is liable to worsen.

The advent of automated vehicles offers unique opportunities to reduce congestion, as control of the route is removed from human hands. Improving vehicle routes, according to some objective function, could become a new and powerful control action that can be taken. The dynamic vehicle routing problem has been concerned with this for some time \cite{Pillac2013, Smith1984, Smith1980}, but there are few clearly defined decentralised and adaptive control strategies for vehicles.

In a vehicular network the routing problem is to find strategies to assign routes to vehicles in order to minimise their travel time and reduce congestion on the network. Strategies can be local, if they only rely on information available to the vehicle in a local neighbourhood of the network, or global if vehicles are assumed to know the state of the entire network. The simplest and most obvious global strategy is Dijkstra's algorithm \cite{Dijkstra1959}, whereby each vehicle takes the shortest path between origin and destination, and its extensions (see for example \cite{Hart1968, Delling2009, Sanders2005, Geisberger2008}). All of these modern routing algorithms enable extremely fast calculation of shortest paths (down to milliseconds), but all of them require some pre-processing of the traffic network and storage of data that is produced from this pre-processing. They are excellent for large static networks, but when a network is dynamic, such approaches are unpractical for large networks, and require a high communications overhead when computed by a central controller and communicated to every vehicle.

Developments have been made for truly novel routing strategies. For instance in \cite{Bell2009, Bell2012}, algorithms were developed for vehicle navigation, which aim to provide highly reliable travel time estimations. The algorithm, named 'Hyperstar', plans routes based on the probability of encountering a delay and needing to divert along a different path. In \cite{Lim2012} a distributed algorithm is presented which can calculate the socially optimum set of traffic flows, given a set of vehicle trips, and then route vehicles probabilistically using those optimum flows.

The aim of this paper is to introduce and explore a simpler, yet effective, local routing algorithm based on using not only shortest path information but also some knowledge of localised congestion on different roads. The idea is that as a vehicle approaches a junction, it is assigned the next road on its route by evaluating a cost function composed of two terms: one related to distance from their destination, the other on the measured congestion level of roads out of that junction. The approach is inspired by coverage control strategies in control theory \cite{Schwager2009} and is compared to other routing strategies on a set of representative examples. We find that, when compared to more traditional strategies, our approach guarantees a lower level of congestion and minimal delays. In particular, we uncover a subtle relationship between the mean travel time, the level of congestion and the weight in the cost function determining the relative dominance of the distance term over the congestion term. This seems to suggest that optimal tuning of the routing control parameters is possible depending on the structural properties of the road network and the car density on it.

% PROBLEM FORMULATION

\section{Coverage Based Routing Algorithm}
\begin{figure}
	\centering
	\subfloat[Coverage based routing\label{fig:coveragebasedrouting}]{\includegraphics[width=0.5\textwidth]{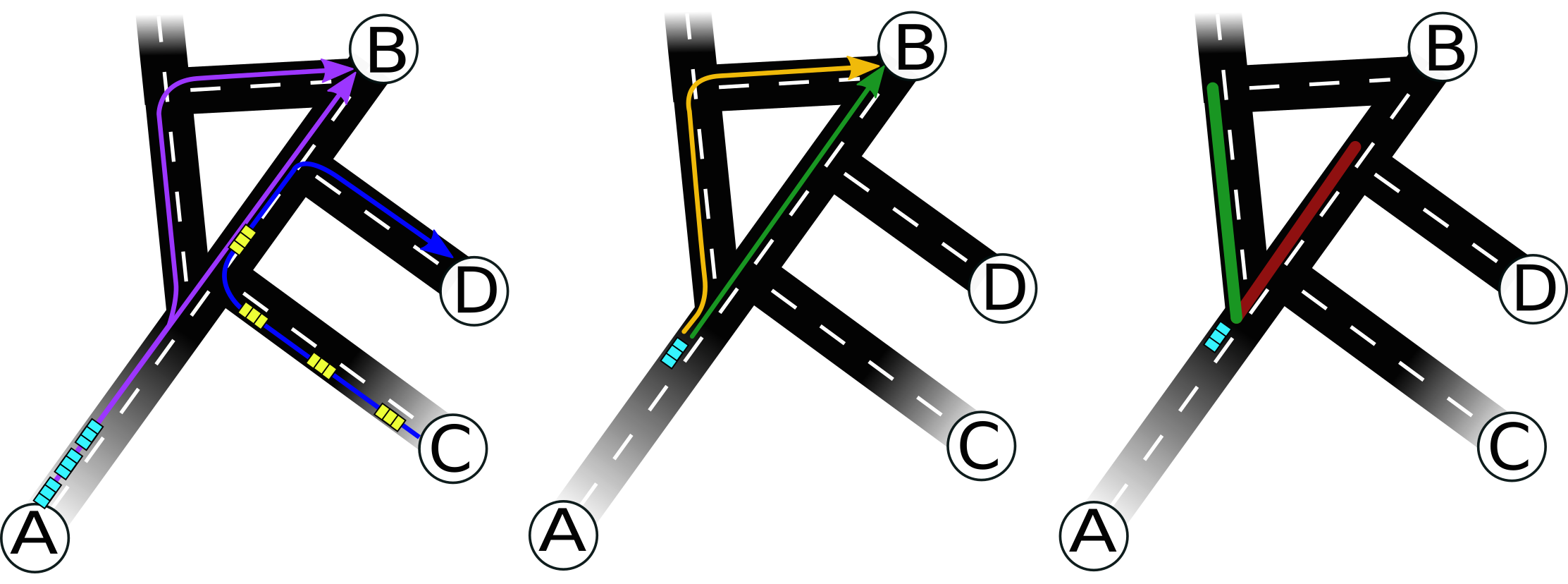}}\\
	\subfloat[Distance cost function\label{fig:theta_definition}]{\includegraphics[width=0.33\textwidth]{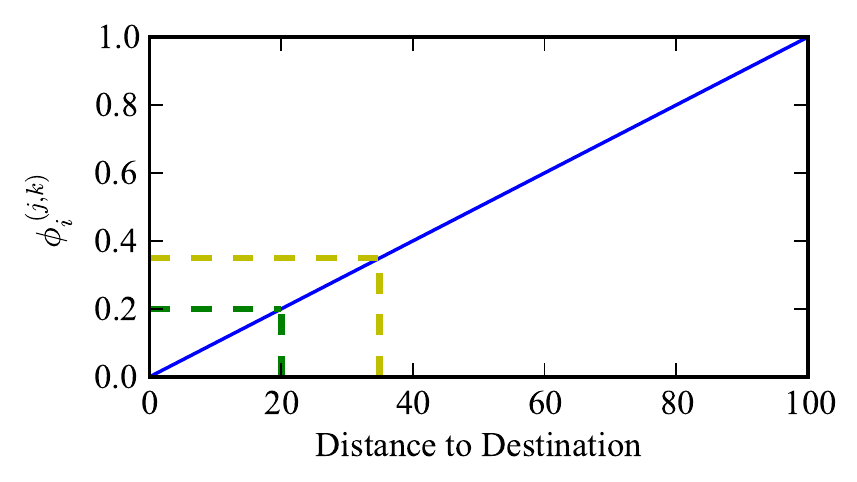}}\\
	\subfloat[Congestion cost function\label{fig:rho_definition}]{\includegraphics[width=0.33\textwidth]{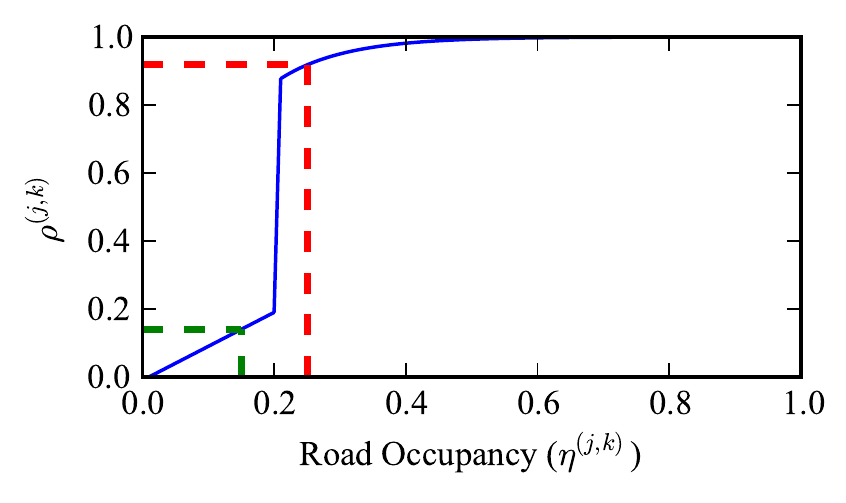}}
	\caption{(a) The principle behind coverage based routing. A stream of cars want to travel between A and B, whilst another group of cars want to travel between C and D. Cars travelling between A and B could take one of two possible routes to get there (far left image). As a car approaches a possible turning point it evaluates the distance of the journey along each potential route (centre image), the yellow arrow indicates the longer route, and a respective cost function is shown in (b), for some arbitrary distances. The car also evaluates the level of congestion on roads it could take (far right image). The red line indicates the road of high occupancy, and the green line indicates the road of low occupancy. The respective cost function is shown in (c). In this example, one route is shorter, but has higher congestion. The car will evaluate the two cost functions and combine them to pick the best route.\label{fig:coveragebasedrouting_whole}}
\end{figure}

Coverage control is a method for controlling distributed mobile sensor networks, in order that they can deploy mobile sensors and provide optimal coverage of a sensory function within a bounded region \cite{Schwager}.
In order to maximise the flow in a traffic network, we hypothesize that it would be preferable to distribute vehicles as evenly as possible along roads, in order to reduce congestion while keeping their travel time minimal. This hypothesis is supported by the traffic flow model proposed by Lighthill and Whitham \cite{Lighthill1955}, which relates traffic density (the number of cars per km of road) to traffic flow (the number of cars passing a particular point per unit time), as well as by further approaches which have used congestion avoidance as a factor for calculating optimal routes for vehicles \cite{Smitha2012, Aslam2012a}. In our routing algorithm, vehicles attempt to balance two sensory functions, one which attracts them towards their destination, and another which repels them from roads already heavily occupied by other vehicles (see Figure \ref{fig:coveragebasedrouting_whole}).
\subsection{Description of Routing Algorithm}
Consider a set of $N$ vehicles, $\mathbf{I} = \{1,2,...,N\}$, in a bounded region $\mathbf{Q} \subset \mathbb{R}^2$. The position of the $i$-th vehicle is denoted by the vector $\mathbf{p_i} \in \mathbf{Q}, i\in I$. The intended final destination of the $i$-th vehicle is denoted by some point $\mathbf{d_i} \in \mathbf{Q}$. Vehicles are located on a road network  $\mathbf{G} = \{\mathbf{V}, \mathbf{E}\}$ contained in $\mathbf{Q}$, where $\mathbf{V} = \{1,2,...,m\}$ is the set of $m$ vertices which represent junctions, and $\mathbf{E}$ is the set of edges which represent the roads between junctions. The $j$-th junction is denoted $v_j$, $j\in \mathbf{V}$. An arbitrary road is denoted $(v_j,v_k) \in \mathbf{E}$, where $v_j$ and $v_k$ are the parent and child vertices (junctions) of the edge (road) respectively. Each road $(v_j, v_k)$ has the following attributes:
\begin{enumerate}
\item A load [$L^{(j,k)}(t)$] - modelling the number of vehicles currently using the road  $(v_j,v_k)$
\item A capacity [$C^{(j,k)}$] - the maximum number of vehicles that can fit onto road $(v_j,v_k)$
\item An occupancy [$\eta^{(j,k)}(t) = \frac{L^{(j,k)}(t)}{C^{(j,k)}}$] - the percentage of space on road $(v_j,v_k)$ occupied by vehicles
\end{enumerate}

When a car, say the $i$-th vehicle, reaches a junction $v_j$, it can access two types of sensory functions $\phi_i^{(j,k)}$ and $\rho^{(j,k)}$, for every road $(v_j,v_k) \in \mathbf{E_j}$, where $\mathbf{E_j}$ is a set of all roads connected to junction $v_j$.
The first sensory function, $\phi^{(j,k)}_{i}$, is related to the estimated distance that car $i$ will travel if it takes a particular road $(v_j,v_k) \in \mathbf{E_j}$ on the way to its destination:
\begin{equation}
\begin{split}
\phi^{(j,k)}_{i} = \phi(d_i,v_j,v_k),  \quad 0 \le \phi^{(j,k)}_{i} \le 1 \\
\label{eq:phi}
\end{split}
\end{equation}
The second sensory function, $\rho^{(j,k)}$, is related to the occupancy $\eta^{(j,k)}(t)$ of road $(v_j,v_k)\in \mathbf{E_j}$ at time $t$:
\begin{equation}
\begin{split}
\rho^{(j,k)} = \rho(\eta^{(j,k)}(t)), \quad 0 \le \rho^{(j,k)} \le 1 \\
\label{eq:rho}
\end{split}
\end{equation}

Expressions (\ref{eq:phi}) and (\ref{eq:rho}) are combined via a tuning parameter $\alpha\in[0,1]$ in order to assign a cost function $J^{(j,k)}_i$ to each road $(v_j, v_k)$ available to car $i$. When car $i$ reaches  junction $v_j$, it computes the cost functions for every road $(v_j,v_k) \in \mathbf{E_j}$ as,
\begin{equation}
\begin{split}
J^{(j,k)}_{i} = \alpha \phi^{(j,k)}_{i} + (1 - \alpha) \rho^{(j,k)}
\label{eq:J}
\end{split}
\end{equation}
Car $i$ at junction $v_j$ will then take the road $(v_j,v_k) \in E_j$ such that
\begin{equation}\label{minj}
\min_{(v_j,v_k) \in \mathbf{E}_j}J^{(j,k)}_{i}
\end{equation}
Table \ref{tab:algorithm} summarises the routing algorithm. Note that by tuning the control parameter $\alpha$ we can make the vehicle more or less sensitive to distance or congestion respectively.
%We will see in what follows that optimal values of $\alpha$ exist given the network structure and the amount of traffic on it (see Section \ref{sec:finding_an_optimal_alpha} for further details).

\begin{table}
	\caption{\label{tab:algorithm}Coverage Based Routing Algorithm}
	\begin{center}
		\begin{tabular}{lcl}
			\midrule
			1: \textbf{if} car $i$ is approaching junction $v_j$ \\
			2: \quad Collect local road occupancy data from a nearby intersection controller \\
			3: \quad \textbf{for} every road $(v_j,v_k) \in \mathbf{E_j}$ \\
			4: \quad\quad Compute the distance based sensory function using (\ref{eq:phi}) \\
			5: \quad\quad Calculate the occupancy based sensory function using (\ref{eq:rho}) \\
			6: \quad\quad Combine to find the overall cost function using (\ref{eq:J}) \\
			7: \quad \textbf{end for} \\
			8: \quad Take the road  $(v_j,v_k)$ with the lowest cost function by solving (\ref{minj})\\
            9: \textbf{end if} \\
			\bottomrule
		\end{tabular}
	\end{center}
\end{table}

\subsection{Choice of Sensory Functions}
The sensory function $\phi^{(j,k)}_{i}$ has been chosen as follows
\begin{equation*}
\phi^{(j,k)}_{i} = \frac{\mathcal{D}(v_k,d_i) + \mathcal{D}(v_j,v_k)}{\max_{l \in V}\mathcal{D}(v_l,d_i)+ \max_{(s,t) \in E}\mathcal{D}(v_s,v_t)}
\end{equation*}
where $\mathcal{D}(v_k,d_i)$ is the network distance (shortest path) between junction $v_k$ and destination $d_i$, and $\mathcal{D}(v_j,v_k)$ is the distance of road $(v_j,v_k)$. Additionally, $\max_{l \in V}\mathcal{D}(v_l,d_i)$ denotes the largest network distance between the arbitrary junction $v_l$ and destination $d_i$. By contrast, $\max_{(s,t) \in E}\mathcal{D}(v_s,v_t)$ represents the largest distance of the road in $E$. The basic idea behind this definition is that we want to normalize and quantify the effect of distance when car $i$ takes road $(v_j,v_k)$ and then moves to the destination $d_i$ along the shortest path. To allow for the effect of congestion, the sensory function $\rho^{(j,k)}$ is chosen as
\begin{equation*}
\rho^{(j,k)} =
\begin{cases}
   \eta^{(j,k)}(t), \quad \mbox{if} \quad \eta^{(j,k)}(t) < \eta^{(j,k)}_{\text{crit}} \\
   1 - e^{-\sigma\eta^{(j,k)}(t)}, \quad \mbox{otherwise}
\end{cases} \\
\end{equation*}
where $\eta^{(j,k)}_{\text{crit}}$ is the critical occupancy of road $(v_j,v_k)$, at which the onset of congestion begins, and $\sigma$ is a tuning parameter determining the sensitivity of $\rho^{(j,k)}$ when road $(v_j,v_k)$ goes beyond its critical occupancy.

We were motivated to design the cost function in this way in order to limit the in-flow of vehicles to intersections with long or growing queues, which is observed in traffic flow studies such as \cite{Kerner2011} to reduce the probability of gridlock. A graph of the congestion function $\rho^{(j,k)}$ is shown in Figure \ref{fig:rho_definition}, where the chosen values for the function parameters are $\eta^{(j,k)}_{\text{crit}} = 0.2$ and $\sigma = 10$.

The choice of $\sigma$ is intended to emphasise the strong cost of going above $\eta^{(j,k)}_{\text{crit}}$, with an exponential decay used so that we maintain some distinction between the cost functions for a set of roads which have all gone above their critical occupancy. It is worth pointing out that the above sensory functions can be selected differently and be made dependent on diverse routing factors such as travel time, congestion level and road condition.

%\begin{figure}
%	\centering
%	\includegraphics[scale=1]{rho_definition}
%	\caption{Relationship between sensory function $\rho^{(j,k)}$ and occupancy $\eta^{(j,k)}$.}
%	\label{fig:rho_definition}
%\end{figure}

% A NOVEL ALGORITHM BASED ON COVERAGE CONTROL

% NUMERICAL ANALYSIS & SUMO MODELLING
\section{Numerical Validation}
To validate the strategy we carried out numerical simulations on some representative examples, first in MATLAB \cite{MATLAB} and then using Simulation of Urban Mobility (SUMO) \cite{Krajzewicz2002}, a widely recognised micro agent simulatior for traffic networks. We compare the performance of the coverage based routing algorithm with others based only on shortest path computation. In particular, our algorithm is contrasted with Dijkstra's and a modified Dijkstra algorithm discussed in \cite{Manfredi} in the context of communication networks that, when more than one shortest path is available, takes the one with the next shortest queue of data packets.

\subsection{MATLAB Implementation}

\begin{figure}
	\centering
	\subfloat[\label{fig:lambda_vs_routing_comparison_matlab}]{\includegraphics[width=0.4\textwidth]{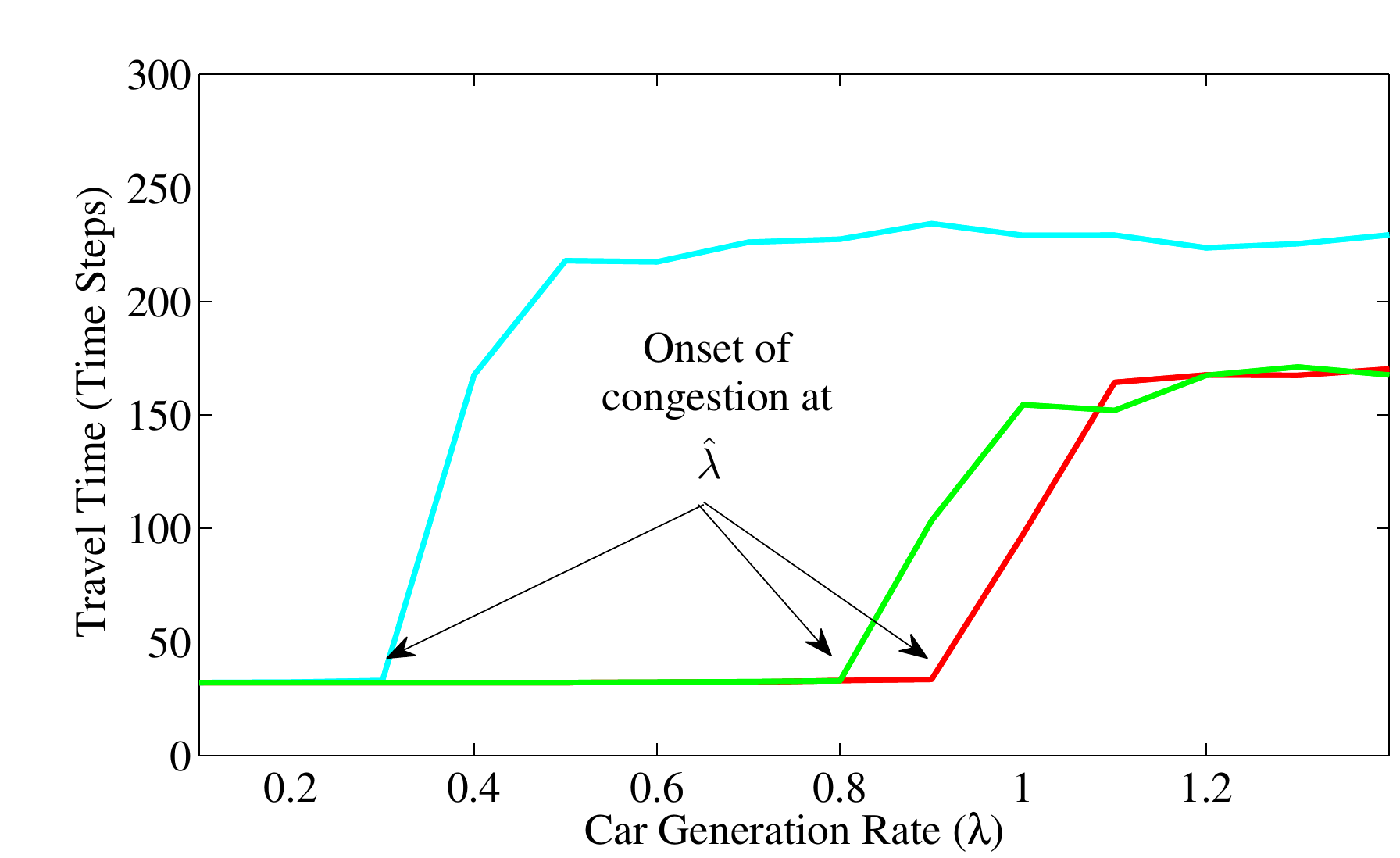}}\\
	\subfloat[\label{fig:optalphamatlab}]{\includegraphics[width=0.45\textwidth]{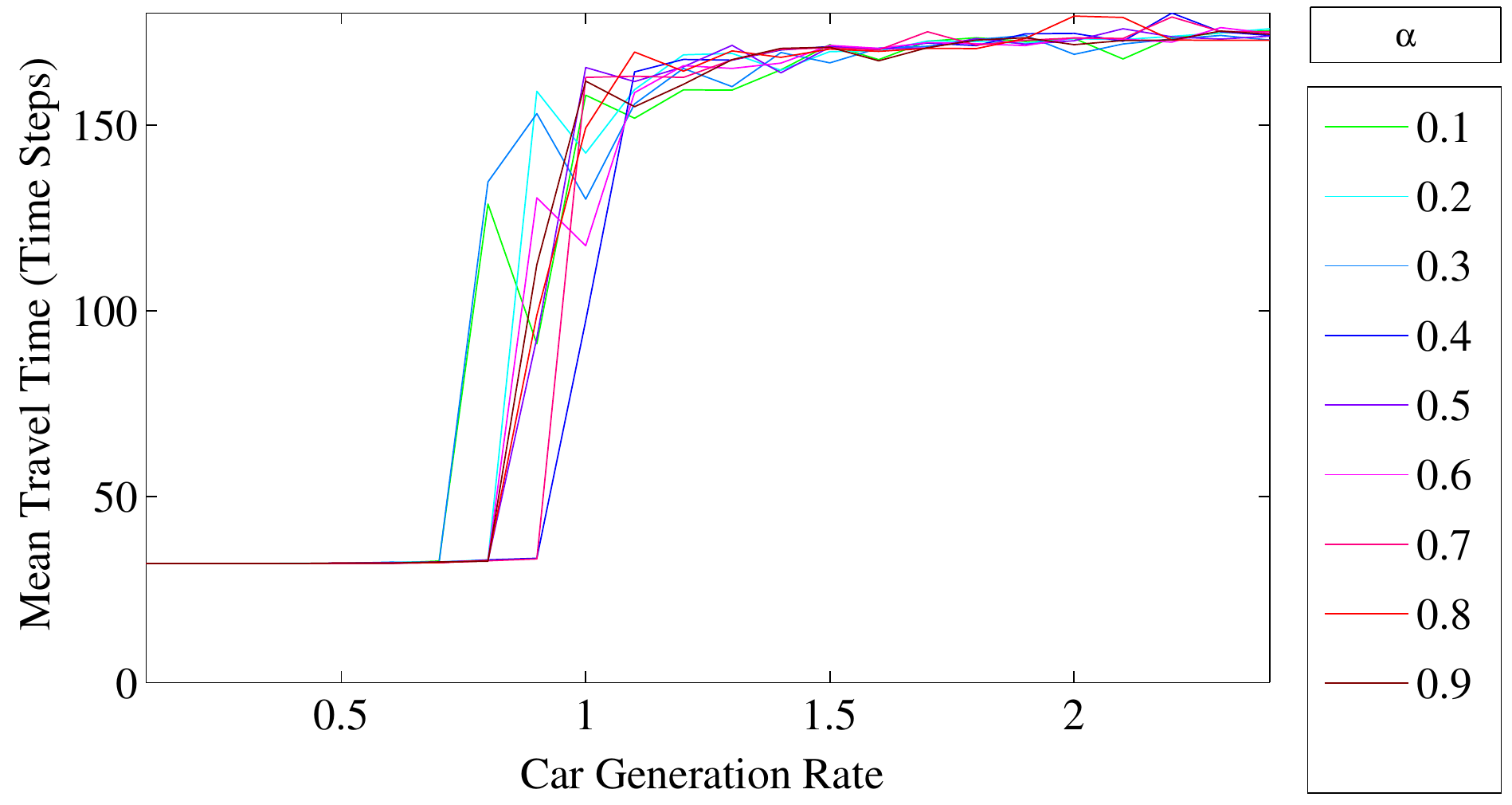}}
	\caption{Results from modelling in Matlab. (a) Mean travel time plotted against car generation rate for coverage based routing when $\alpha = 0.4$ (red), shortest path routing using Dijkstra (cyan) and a modified shortest path algorithm (green). (b) Effect of varying $\alpha$ on mean travel time for coverage based routing.}
    	\label{fig:matlab_results}
\end{figure}

In our MATLAB implementation, each vehicle calculates a new position at every time step, based on the average speed of the road it is travelling on (average speed is estimated using an occupancy-flow relationship from \cite{Hall1986}). When a car reaches a junction, it recalculates the optimum route using the routing algorithm being used. Cars are generated in the network according to a Poisson process of rate $\lambda$, we shall term the car generation rate. We denote by $\hat{\lambda}$ the maximal value of the car generation rate at which congestion is observed to occur (see Figure \ref{fig:lambda_vs_routing_comparison_matlab}).

Each routing algorithm was tested for increasing car generation rates in order to see the effect on the routing algorithm performance. The behaviour of the modified Dijkstra algorithm was very similar to coverage based routing when $\alpha=1$. Given the correct choice of the tuning parameter $\alpha$, coverage based routing outperformed the other routing algorithms (see Figure \ref{fig:matlab_results} where we show results for a 3x3 grid network, where all vehicles make a trip between the bottom-left and top-right corners of the grid). To further test this observation in a more computationally effective way and on larger networks, we move next to verify the algorithm in SUMO.

\subsection{SUMO Verification}
To investigate the performance of the algorithm and confirm the importance of tuning the parameter $\alpha$, we used SUMO. Using the SUMO API (TraCI), several Python modules were produced in order to allow vehicles to be rerouted dynamically at each junction, according to the selected routing algorithm.

For all simulations, journeys were generated randomly using the 'randomTrips.py' script (included within the SUMO tools library) and traffic lights were omitted. In these simulations the car generation rate refers to the number of cars entering the network at each time-step, which is constant and not a poisson process.

Mean delay was measured as being more informative than the travel time in the scenarios of interest. Indeed trips between different nodes had different expected travel times, whereas delay was directly comparable between all trips. The mean delay for a journey between any Origin-Destination (O-D) pair, is calculated as the difference between the fastest possible travel time for any vehicle between the O-D pair, and the actual experienced travel time for a vehicle making the same trip during a simulation.

The key concepts we wished to investigate were:

\begin{itemize}
\item The existence of an optimal region of $\alpha$, which maximises network capacity and reduces congestion.
\item The performance of coverage based routing against routing based on the shortest path only.
\item The effect of different road network structures on the performance of the algorithm
\end{itemize}

\subsubsection{Network Topologies \label{sec:net_topologies}}
We studied the performance of coverage based routing on a selection of network topologies, in order to better understand how topology affected its performance. All roads in the networks are given the same speed limit, so that calculation of the shortest path by distance is equivalent to calculation of the shortest path by travel time (where travel-time is not updated according to real-time traffic data) and there are no 'motorway' shortcuts. The networks are undirected, so there are no one-way streets.

\begin{figure}
	\centering
	\subfloat[Random network\label{fig:random_1_net}]{\includegraphics[width=0.18\textwidth]{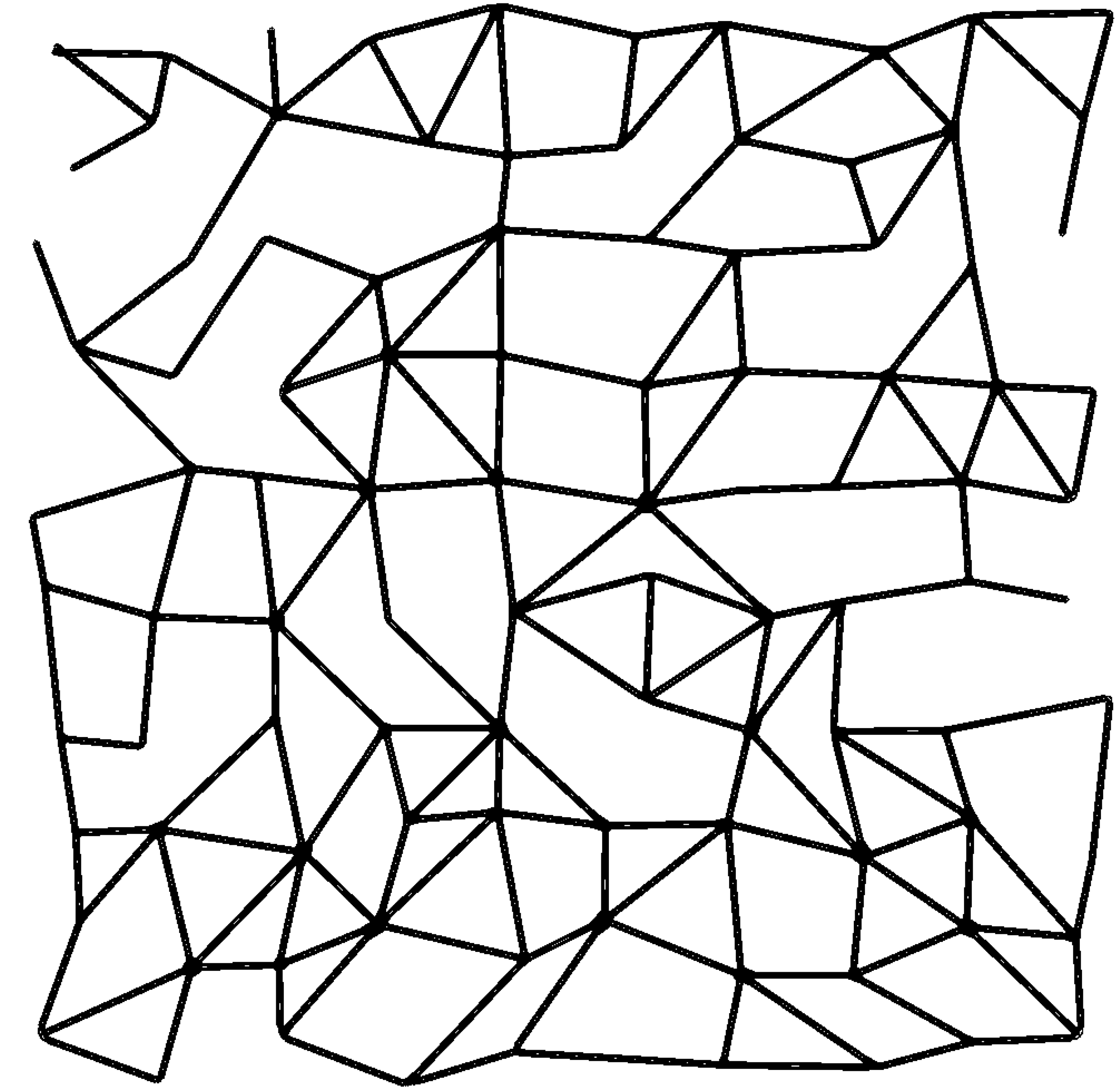}} \\
	\subfloat[Spiderweb network\label{fig:spider_1_net}]{\includegraphics[width=0.18\textwidth]{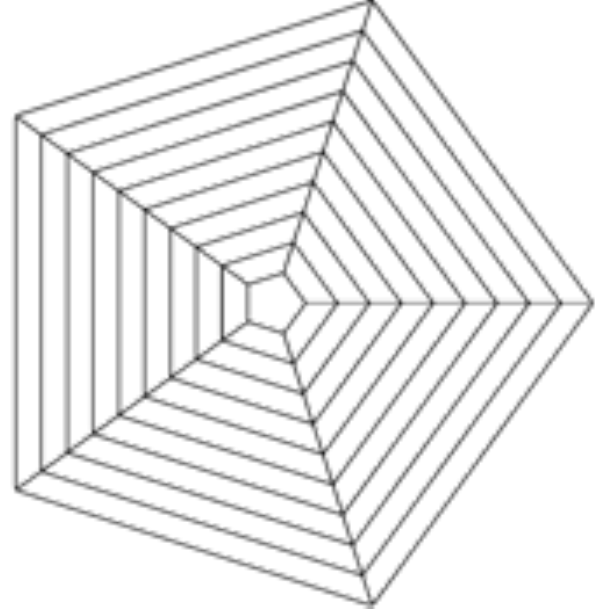}} \quad
	\subfloat[Scale-free network\label{fig:scale_free_1_net}]{\includegraphics[width=0.18\textwidth]{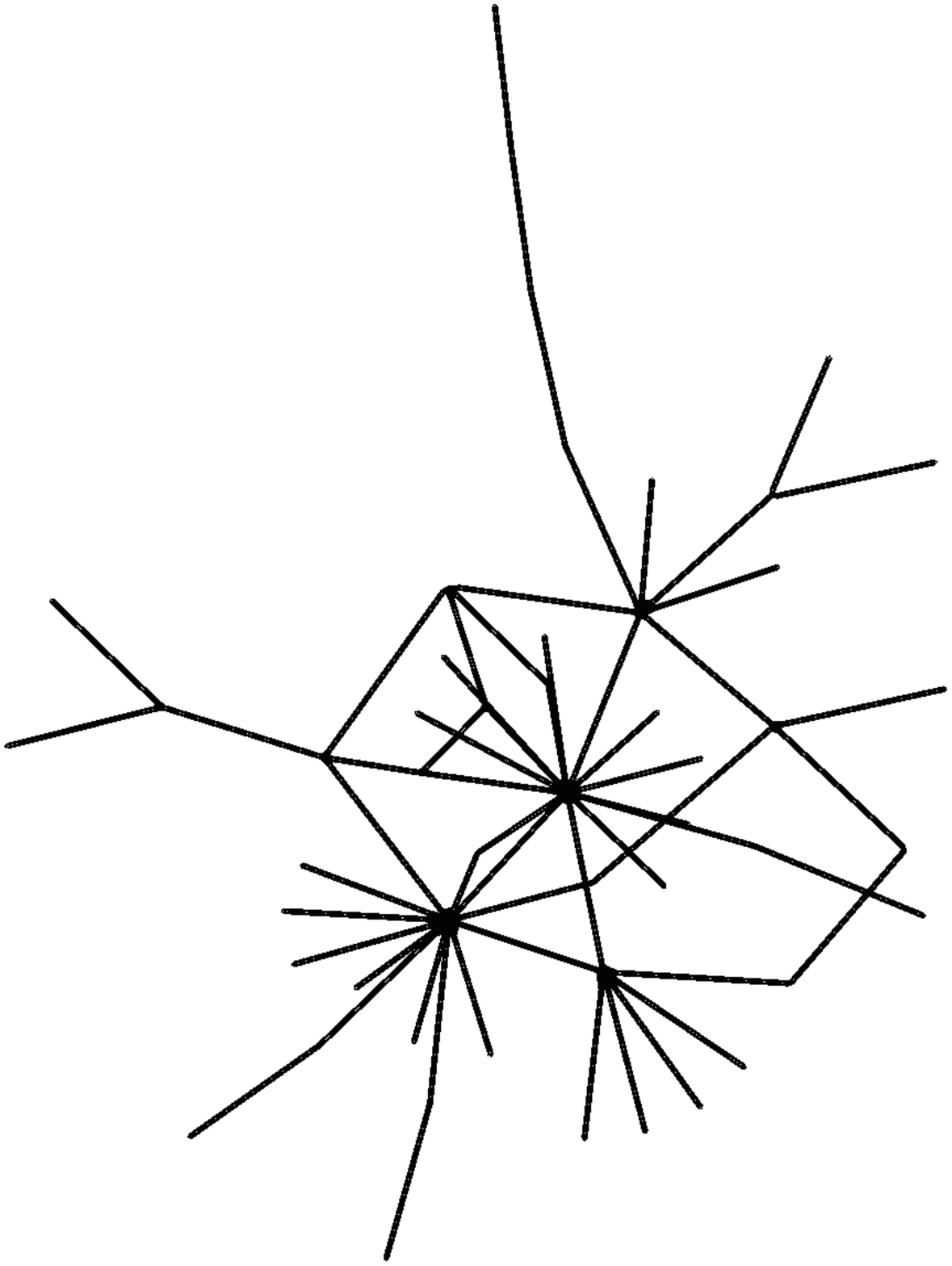}}
        \caption{Network topologies used in the SUMO simulations\label{fig:topologies}}
\end{figure}

Tests run using SUMO included a 5x5 and 10x10 node grid network, a spiderweb network, a random network, and a scale-free network (see Figure \ref{fig:topologies} and Table \ref{tab:network_properties} for their structural properties). 

The two grid networks analysed are inspired by road layouts in places such as Manhattan Island in New York City (Google Map - \url{http://tinyurl.com/kz32ut8}). Any vehicle wanting to travel diagonally across the network can choose among a large selection of shortest paths, thereby favouring a routing algorithm which considers all the possible paths.
The random network (Figure \ref{fig:random_1_net}) was generated by rewiring the 10x10 grid network, and links were generated using a similar approach to the Erd\"{o}s-R\'{e}nyi Planar Graph model in \cite{Masucci2009}. The random network is unlikely to offer many paths of equal distance between destinations, and so vehicles must balance the gains from avoiding congestion, with the cost of taking a longer path.
The spiderweb network (Figure \ref{fig:spider_1_net}) is a collection of ring roads of increasing size, connected by inward 'spokes'. This is a simplification of the road structure found in cities such as Beijing (Google Map - \url{http://tinyurl.com/ojyexwk}). In this road layout many vehicles routing using only the shortest path will attempt to use the central ring as a shortcut between destinations.
Finally, the scale-free network is inspired by the evidence that a large number of urban road networks are scale-free networks \cite{Jiang2007}. 
%
%We found that the network topology had a significant impact on both the optimal value of the tuning parameter $\alpha$, and on the overall performance of both shortest path and coverage based routing (these findings are detailed in Sections \ref{sec:finding_an_optimal_alpha} and \ref{sec:comparison_of_routing_algorithms}).
%
\subsubsection{An Optimal Range of $\alpha$ \label{sec:finding_an_optimal_alpha}}
\begin{figure}
	\centering
	\subfloat[Mean delay against $\alpha$ and car generation rate (blue = lower delay, red = higher delay). Dashed lines indicate the sections shown in Figure \ref{fig:grid_1_amd}.\label{fig:grid_1_avcgrd}]{\includegraphics[width=0.35\textwidth]{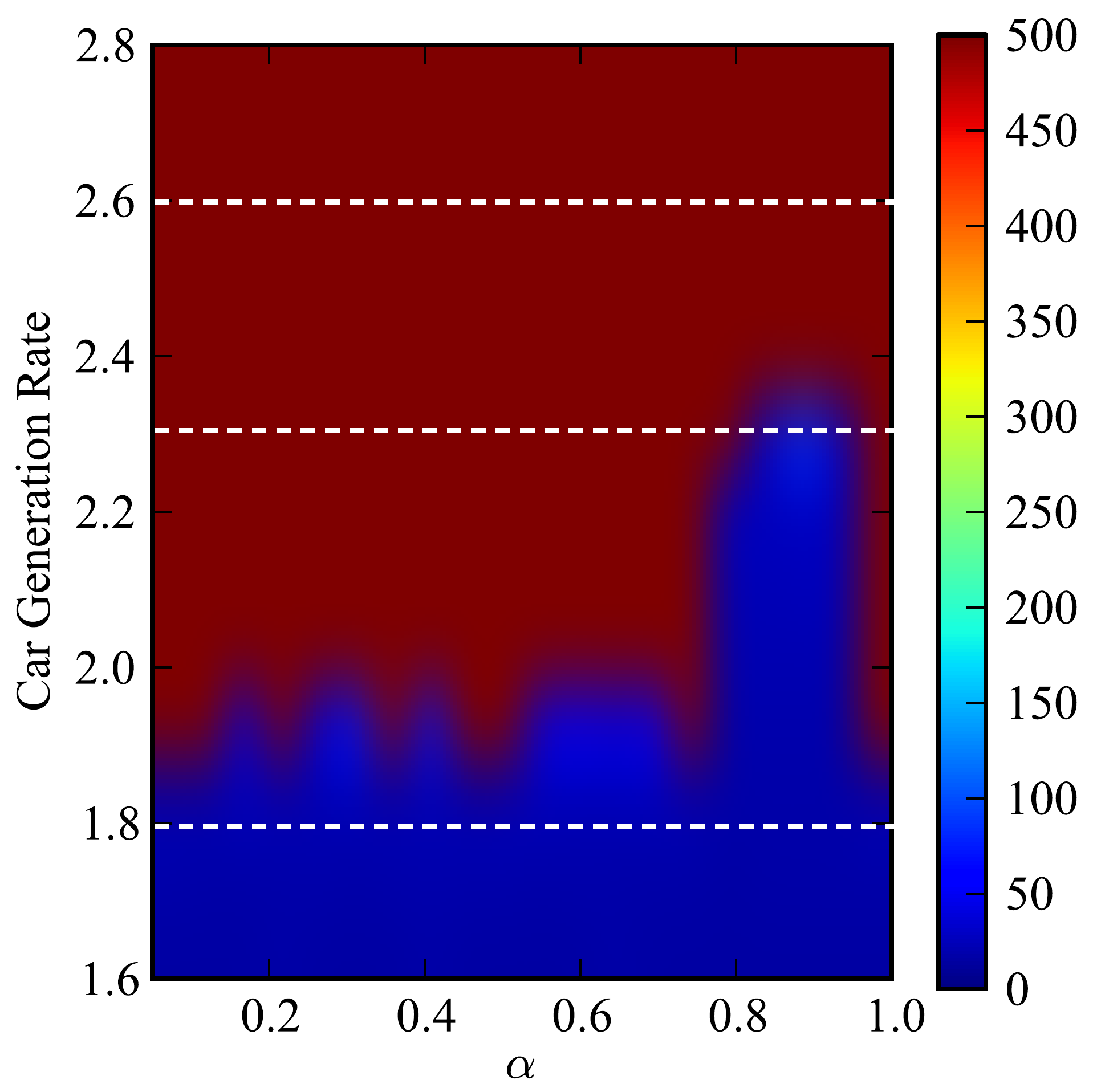}} \\
		\subfloat[Cut-throughs of Figure \ref{fig:grid_1_avcgrd} when car generation rate is 1.8 (blue), 2.3 (green), and 2.6 (red). \label{fig:grid_1_amd}]{\includegraphics[width=0.35\textwidth]{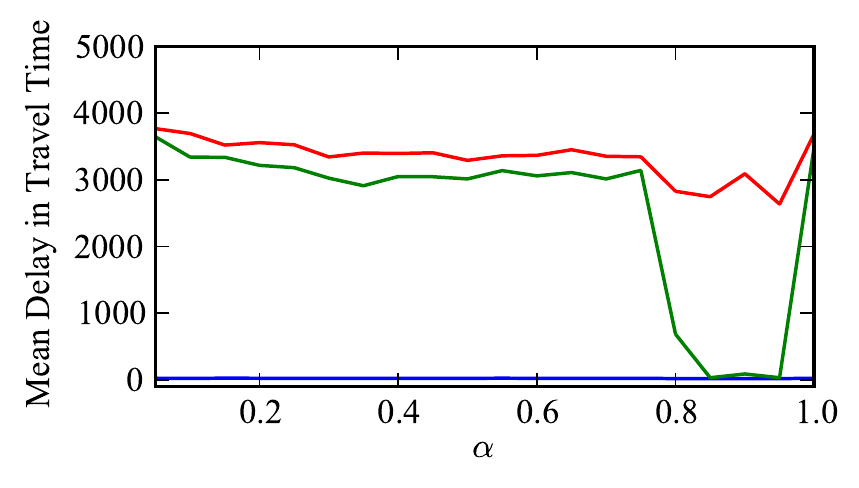}} \\
	\caption{5x5 Grid: Mean delay as a function of $\alpha$ and $\lambda$}
	\label{fig:grid_1_optalpha}
\end{figure}
\begin{figure}
	\centering
	\subfloat[Mean delay against $\alpha$ and car generation rate (blue = lower delay, red = higher delay). Dashed lines indicate the sections shown in Figure \ref{fig:grid_5_amd}. \label{fig:grid_5_avcgrd}]{\includegraphics[width=0.3\textwidth]{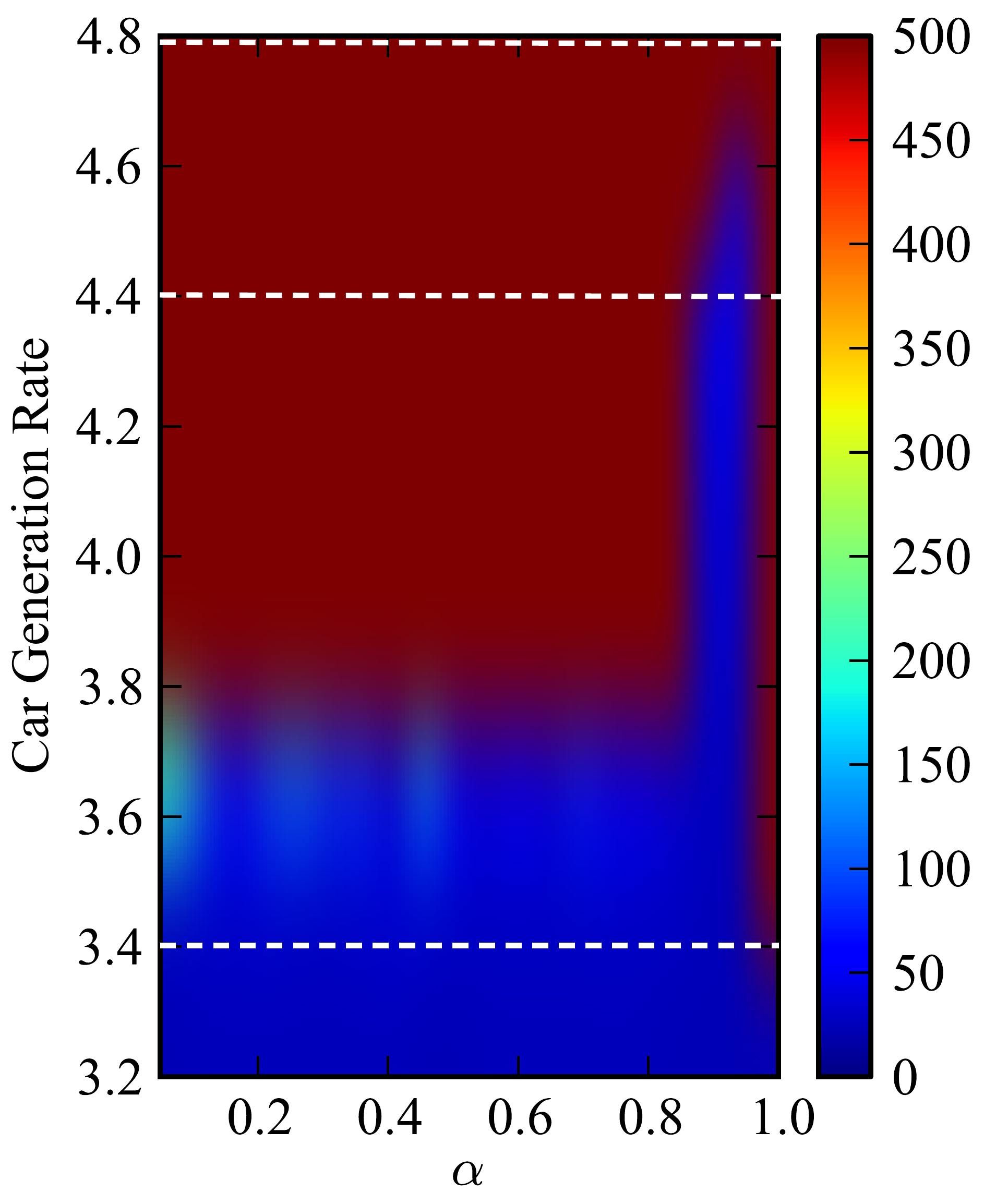}} \\
		\subfloat[Cut-throughs of Figure \ref{fig:grid_1_avcgrd} when car generation rate is 3.4 (blue), 4.4 (green), and 4.8 (red).\label{fig:grid_5_amd}]{\includegraphics[width=0.35\textwidth]{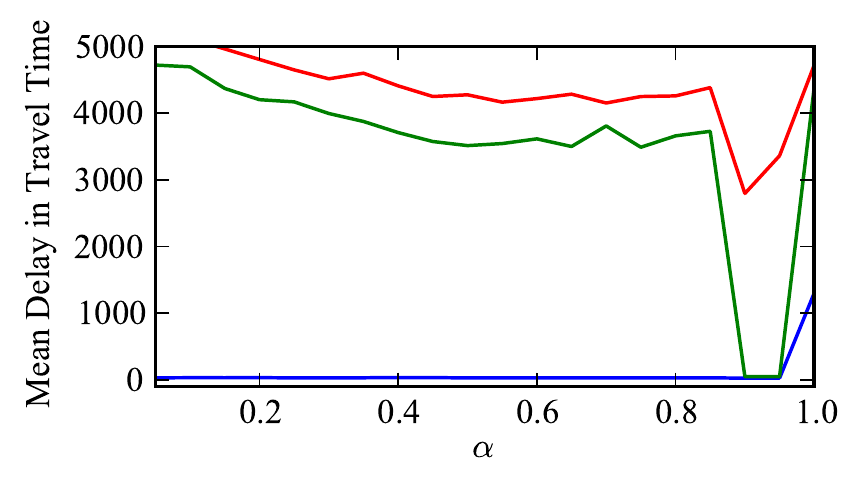}} \\
	\caption{10x10 Grid: Mean delay as a function of $\alpha$ and $\lambda$}
	\label{fig:grid_5_optalpha}
\end{figure}
\begin{figure}
	\centering
	\subfloat[Mean delay against $\alpha$ and car generation rate (blue = lower delay, red = higher delay). Dashed lines indicate the sections shown in Figure \ref{fig:ERPG_1_amd}.\label{fig:ERPG_1_avcgrd}]{\includegraphics[width=0.3\textwidth]{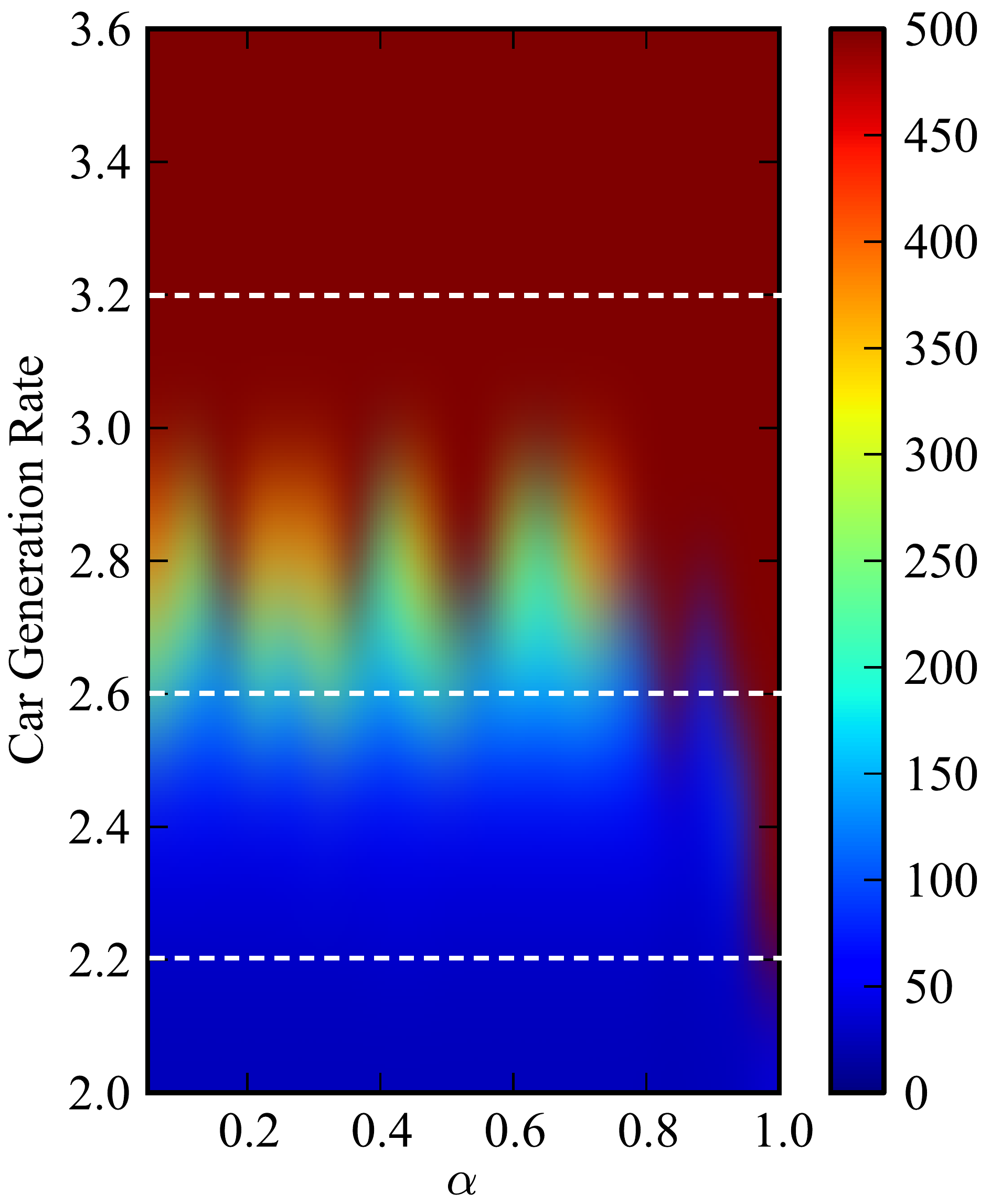}} \\
		\subfloat[Cut-throughs of Figure \ref{fig:grid_1_avcgrd} when car generation rate is 2.2 (blue), 2.6 (green), and 3.2 (red).\label{fig:ERPG_1_amd}]{\includegraphics[width=0.35\textwidth]{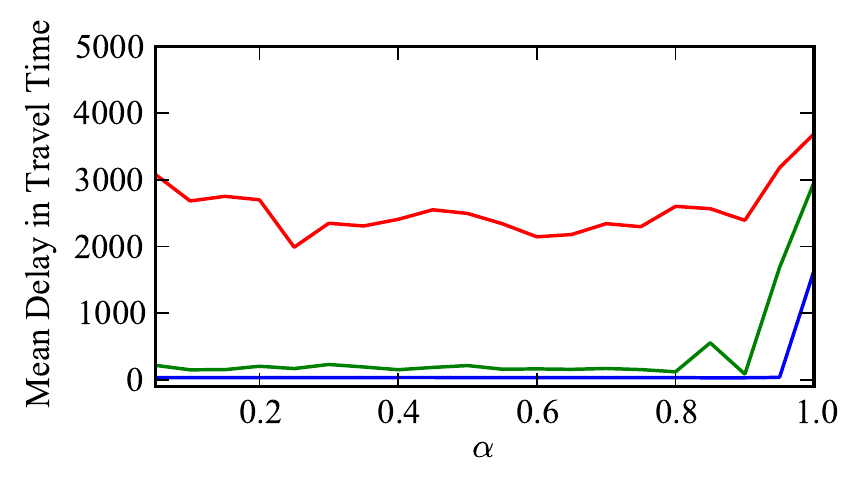}} \\
	\caption{Random: Mean delay as a function of $\alpha$ and $\lambda$}
	\label{fig:ERPG_1_optalpha}
\end{figure}
\begin{figure}[ht]
	\centering
	\subfloat[Mean delay against $\alpha$ and car generation rate (blue = lower delay, maroon = higher delay). Dashed lines indicate the sections shown in Figure \ref{fig:spider_1_amd}. \label{fig:spider_1_avcgrd}]{\includegraphics[width=0.35\textwidth]{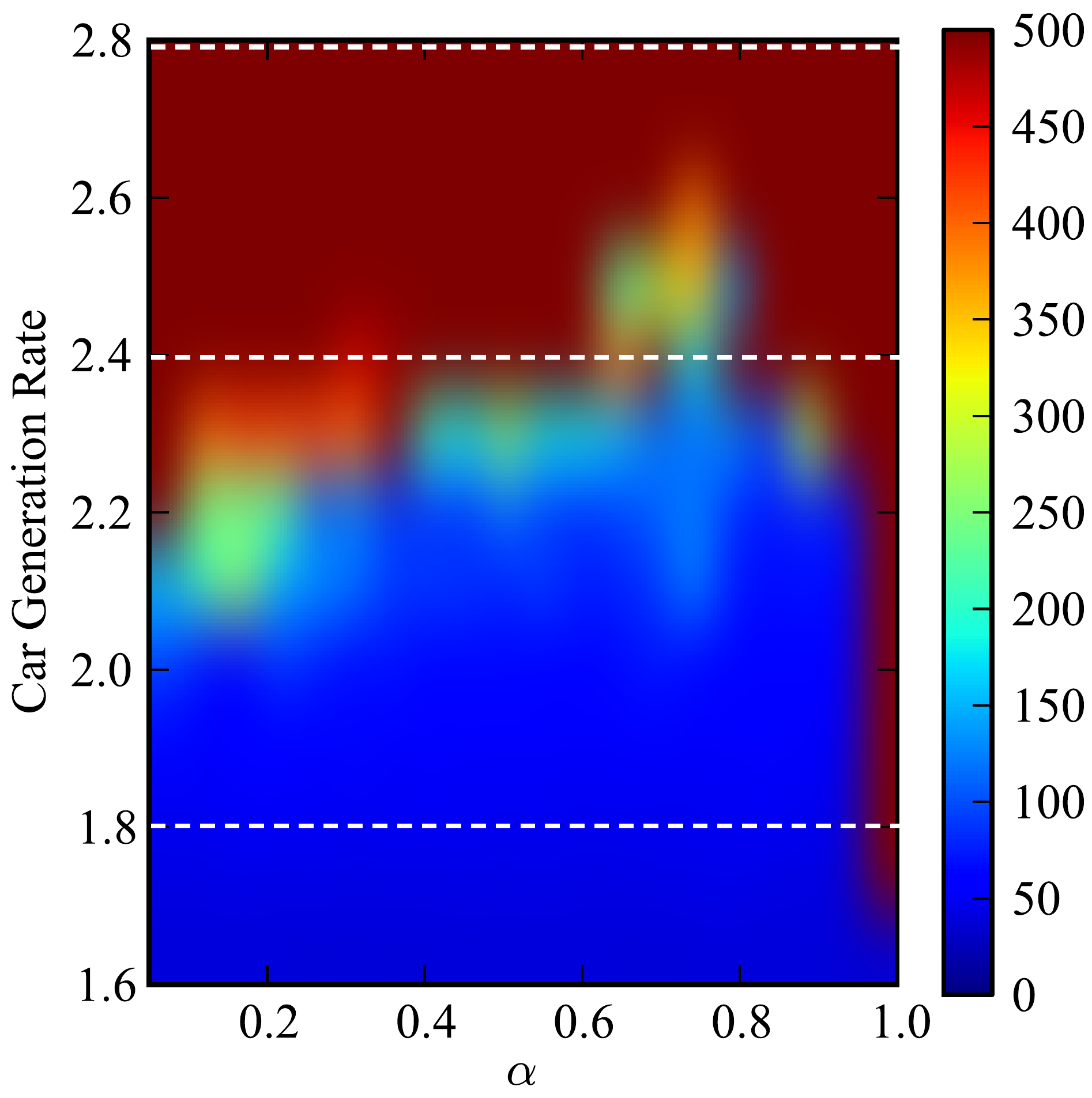}} \\
	\subfloat[Cut-throughs of Figure \ref{fig:spider_1_avcgrd} when car generation rate is 1.8 (blue), 2.4 (green), and 2.8 (red). \label{fig:spider_1_amd}]{\includegraphics[width=0.35\textwidth]{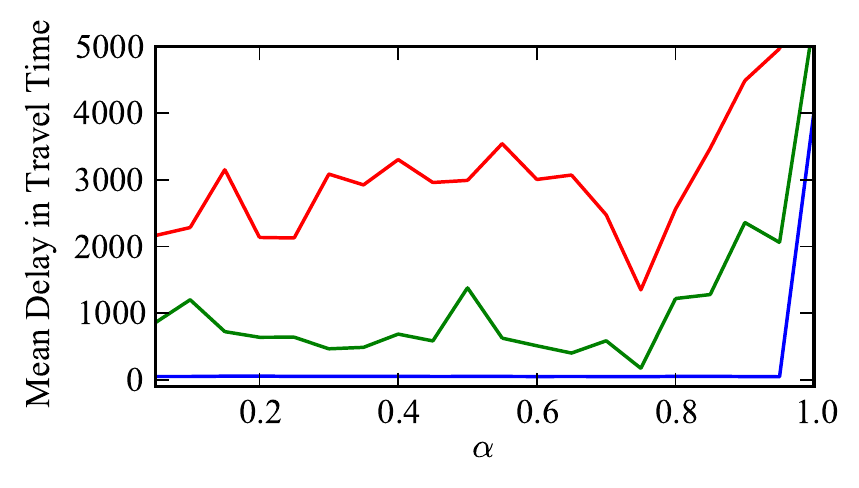}} \\
	\caption{Spiderweb: Mean delay as a function of $\alpha$ and $\lambda$}
	\label{fig:spider_1_optalpha}
\end{figure}
\begin{figure}[t]
	\centering
	\subfloat[Mean delay against $\alpha$ and car generation rate (blue = lower delay, red = higher delay). Dashed lines indicate the sections shown in Figure \ref{fig:scale_free_1_amd}.\label{fig:scale_free_1_avcgrd}]{\includegraphics[width=0.35\textwidth]{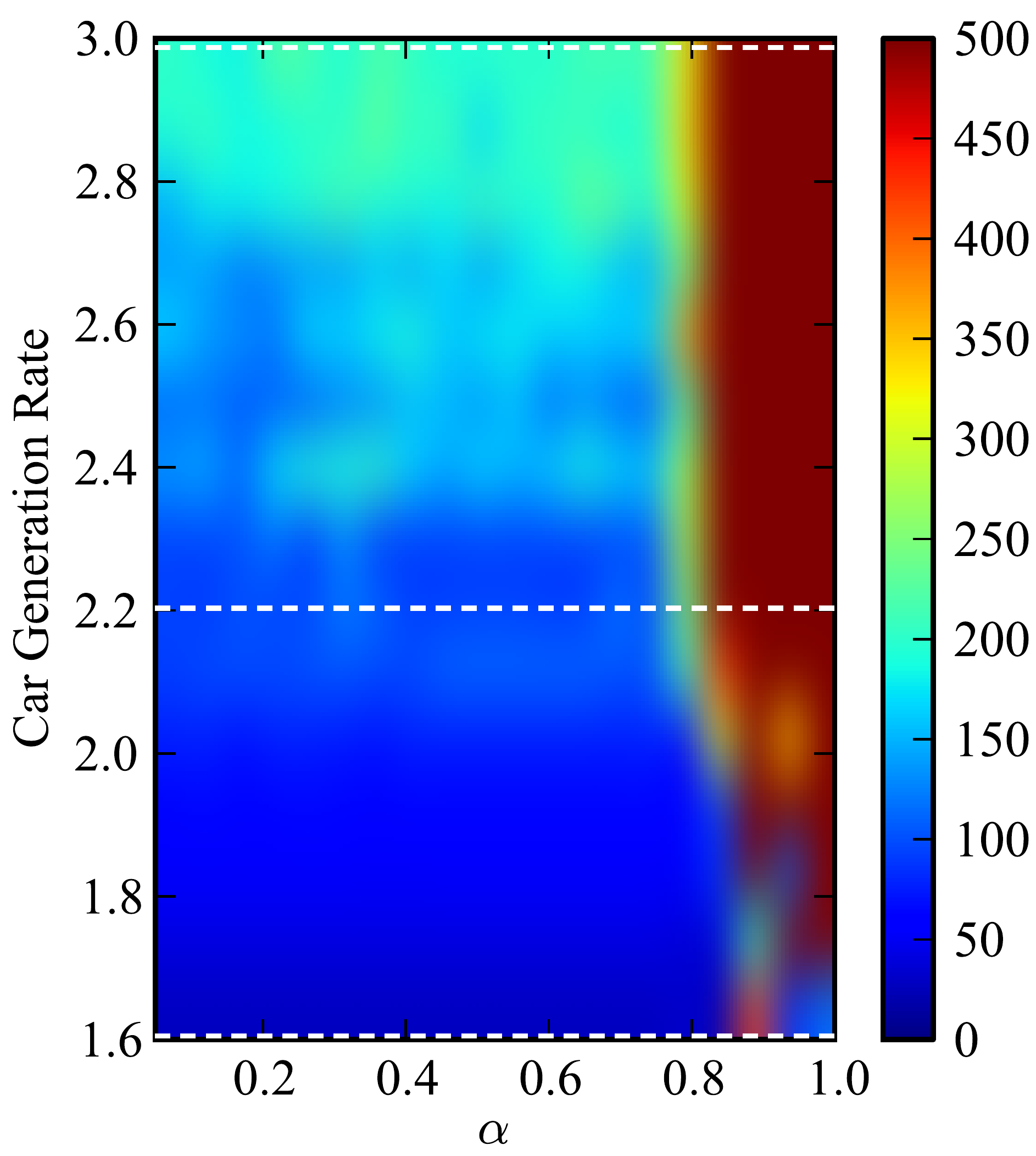}} \\
		\subfloat[Cut-throughs of Figure \ref{fig:scale_free_1_avcgrd} when car generation rate is 1.6 (blue), 2.2 (green), and 3.0 (red). \label{fig:scale_free_1_amd}]{\includegraphics[width=0.35\textwidth]{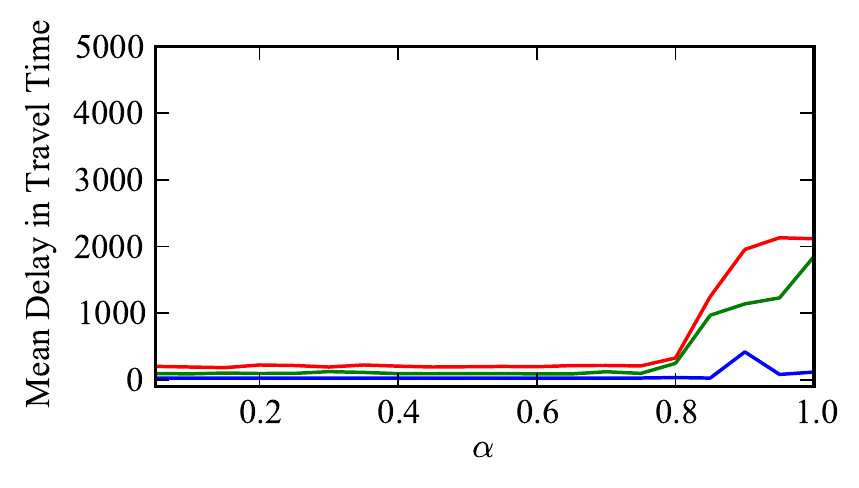}} \\
	\caption{Scale-Free: Mean delay as a function of $\alpha$ and $\lambda$}
	\label{fig:scale_free_1_optalpha}
\end{figure}

Figures~\ref{fig:grid_1_optalpha} to \ref{fig:scale_free_1_optalpha} show the result of approximately 500 simulations for each network topology. Specifically, each point in the diagram is obtained by evaluating the mean delay computed over 3600 time units (1 hour) while the car generation rate and the control parameter $\alpha$ are fixed at a certain value. We observe that the existence of an optimal value for the parameter $\alpha$ is dependent on network topology, and also that when it exists it can be robust to the scale of the network.

We consider that the network becomes congested when an increase in the car generation rate causes an exponential increase in the mean delay (see Table \ref{tab:tableofvalues} for the optimal ranges identified from these plots, associated to the largest value for the car generation rate that the network absorbed before becoming congested  ($\hat{\lambda}$)).

We have capped the mean delay shown at 500 time steps, as although we can observe much higher delays in this model, we are most interested in characterising the onset of congestion. The phase transition to congestion in the network is characterised in part (a) of Figures ~\ref{fig:grid_1_optalpha} to \ref{fig:scale_free_1_optalpha} by the transition between the maroon region (high delay, indicating congestion) and the dark blue region (low delay, indicating free flow).
\begin{table}
	\caption{\label{tab:network_properties}Network Properties}
	\begin{center}
		\begin{tabular}{ccccc}
			\midrule
			\textbf{Network} & \textbf{Nodes} & \textbf{Edges} & \textbf{Mean} & \textbf{Diameter} \\
			\textbf{Topology} & & & \textbf{Degree} & \\
			\midrule
			5x5 Grid & 25 & 40 & 3.2 & 800 \\
			10x10 Grid & 100 & 180 & 3.6 & 1800 \\
			Random & 100 & 180 & 3.6 & 1364   \\
			Spiderweb & 50 & 95 & 3.8 & 2000 \\
			Scale-free & 48 & 58 & 2.4 & 1371 \\ 
			\bottomrule
		\end{tabular}
	\end{center}
\end{table}
\begin{table}
	\caption{\label{tab:tableofvalues} Optimal Range of the Parameter $\alpha$}
	\begin{center}
		\begin{tabular}{clc|c}
			\midrule
			\textbf{Network} & \textbf{Optimal $\alpha$} & \multicolumn{2}{c}{\textbf{Max. Car Gen. Rate}} \\
			 \textbf{Topology} & \textbf{Value(s)} & \multicolumn{2}{c}{\textbf{Before Total Congestion}}\\
			 & & Shortest Path & Coverage Based\\
			\midrule
			5x5 Grid & $0.8$ to $0.95$ & 1.5 & 2.3 (+50\%) \\
			10x10 Grid & 0.9 & 2.0 & 4.4 (+120\%) \\
			Random & 0.25 & 2.0 & 2.8 (+40\%)  \\
			Spiderweb & 0.75 & 1.6 & 2.6 (+60\%) \\
			Scale-free & $\le0.8$ & 1.6 & \emph{Limit not found}\\
			\bottomrule
		\end{tabular}
	\end{center}
\end{table}
The results indicate that there are indeed optimal choices for $\alpha$, which can significantly impact the performance of the coverage based routing algorithm. In all simulations values of $\alpha = 0$ or $1$ performed extremely poorly in comparison to any other intermediate value, confirming that it is best for vehicles to make local choices based on a combination of congestion and shortest path arguments. Setting $\alpha = 0$ represents routing based entirely on avoiding congestion, and so we would expect a poor performance, because the vehicle simply avoids congested roads rather than moving towards its destination. Setting $\alpha = 1$ represents routing based entirely on the shortest path (and in the case of multiple shortest paths, one will be chosen at random). Our results, see Table II, also show a clear dependence of the optimal  range of the control parameter $\alpha$ on the network structure.

In the 5x5 grid network (Figure \ref{fig:grid_1_avcgrd}), the optimal range of $\alpha$ is clearly visible in the region 0.80 to 0.95. Within this range we observe a consistent increase in network capacity indicating that in any application of such a cost based routing strategy, ad hoc tuning of the control parameter could have a huge impact on performance. When the network is scaled up to 10x10 nodes, as shown in Figure \ref{fig:grid_5_avcgrd} we see that the optimal value for $\alpha$ is still 0.9, showing that the choice of $\alpha$ is robust to the effects of scale in a grid network.
In the random network, as shown in Figure \ref{fig:ERPG_1_avcgrd} we find that a unique value of $\alpha$ does not emerge as optimal and several values of $\alpha$ perform equally well, and the same was found in the scale-free network (Figure \ref{fig:scale_free_1_avcgrd}). However, in the spiderweb network (Figure \ref{fig:spider_1_avcgrd}) an optimal value of $\alpha$ emerged at 0.75. This indicates that both the existence of a unique value for an optimal $\alpha$, and its value, are dependent upon the structural properties of the network.

We can conclude that selection of the tuning parameter has a strong influence on the performance of the routing algorithm, and that this may be a unique value depending on the network topology. Even in the random network, it was important to choose an intermediate value of $\alpha$ rather than setting it to $0$ or $1$ for the best performance. This performance difference can be further confirmed by taking vertical cut-throughs in Figures \ref{fig:grid_1_avcgrd} to \ref{fig:scale_free_1_avcgrd} and comparing the threshold value of the car generation rate at which congestion is observed to occur, for different routing algorithms and network structures (see Figures \ref{fig:grid_1_amd} to \ref{fig:scale_free_1_amd}).
\begin{figure*}
	\centering
	\subfloat[10x10 Grid \label{fig:grid_5_rmd}]{\includegraphics[scale=0.85]{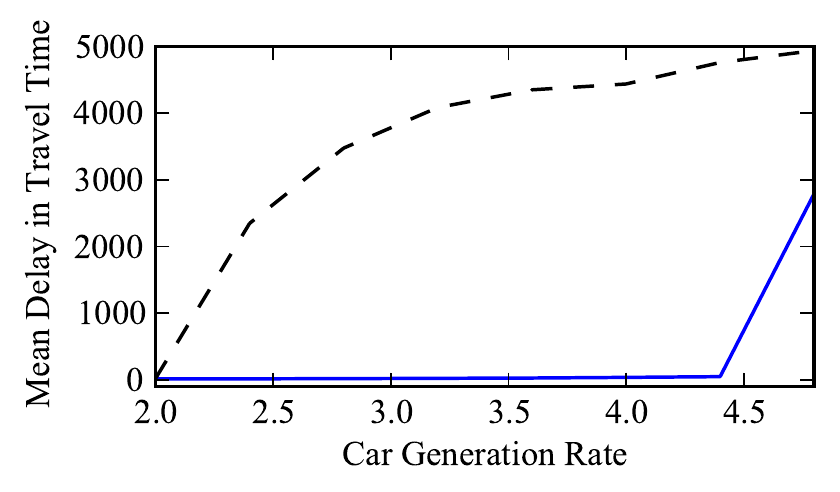}} \quad 
	\subfloat[Random \label{fig:ERPG_1_rmd}]{\includegraphics[scale=0.85]{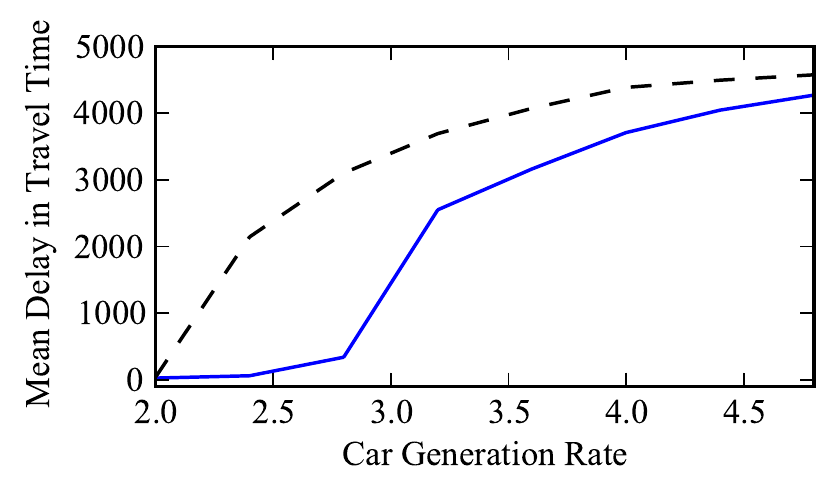}} \\
	\subfloat[Spiderweb \label{fig:spider_1_rmd}]{\includegraphics[scale=0.85]{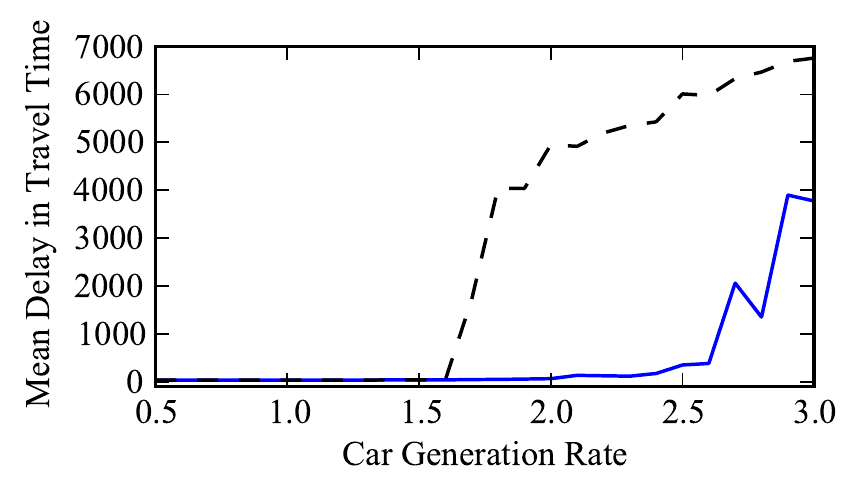}} \quad
	\subfloat[Scale-Free \label{fig:scale_free_1_rmd}]{\includegraphics[scale=0.85]{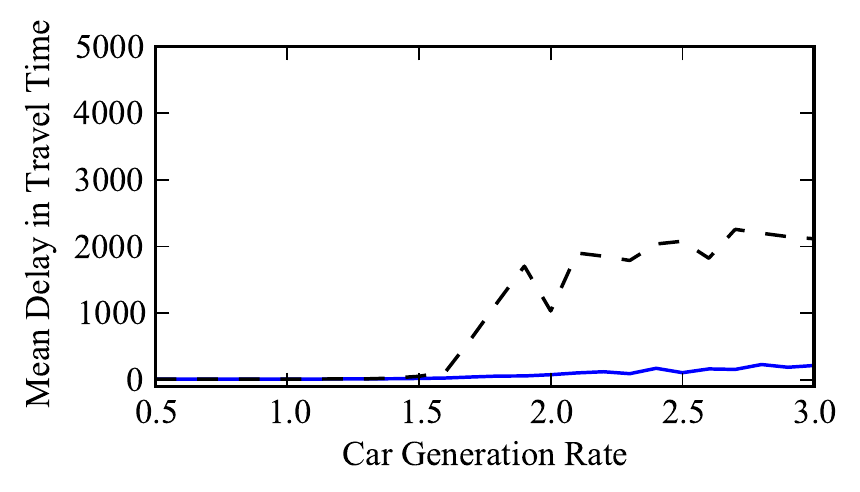}}
	\caption{Comparison of mean delay when routing using the shortest path (dashed black line) and coverage based routing (solid blue line) when using an optimal value of $\alpha$ from Table \ref{tab:tableofvalues}}
\end{figure*}
\subsubsection{Comparison of Routing Algorithms \label{sec:comparison_of_routing_algorithms}}
We have compared our routing algorithm against routing using only a shortest path calculation, as calculated using Dijkstra's algorithm. We find that coverage based routing outperforms routing using only the shortest path, provided the value of $\alpha$ is tuned correctly. 

Figures \ref{fig:grid_5_rmd} to \ref{fig:scale_free_1_rmd} show shortest path based routing and coverage based routing exhibiting similar delays at low car generation rates. However, at some intermediate car generation rates we see shortest path routing unable to keep the network free from congestion, and delays increase by an order of magnitude. In contrast, coverage based routing can continue to maintain much lower delays up to some other, higher, car generation rate. This demonstrates that coverage based routing increases the capacity of the network over routing using only the shortest path. In terms of the increase in maximum car generation rate that coverage based routing exhibits over shortest path routing, we see around a 50\% increase for the 5x5 grid network, a 120\% increase in the 10x10 grid network, a 40\% increase in the random network, a 60\% increase in the spiderweb network, and the absence of a definitive maximum car generation rate in the scale-free network (at least over the range of car generation rates being considered).

The simulations indicate that significant gains in network capacity can be made by using a coverage based approach over shortest path routing, however the size of these gains is dependent upon network topology. To further confirm this finding and allow a better comparison between different network structures, in the final version of this paper we will include simulations on a further set of networks such as a scale-free rewiring of the 10x10 node grid.

\section{Conclusion}
%The conclusion goes here.

The routing strategy we have presented here shows how a simple cost function coupling global information about distances with local road occupancy data can yield improvements over shortest path routing. These improvements are an increase in the capacity of the network, and hence the ability to avoid congestion which leads to delays. We have also shown that in order to optimise this routing strategy a control parameter must be chosen appropriately, and this value appears to depend on the network topology. We envisage that the algorithm could be deployed via a junction infrastructure able to communicate to each vehicle arriving at the junction the best road to take next. We wish to emphasise that the algorithm we present here can be deployed effectively and is easily scalable because of its simplicity. Future work will address the possibility of each car, or junction, becoming able to tune the control parameter $\alpha$ in real time via local adaptive strategies aimed at further minimising congestion and guarantee fairness.

Also, the routing strategy will be tested on real road network topologies, using realistic (asymmetric) loading on the network, including a variety of vehicle types.


\section*{Acknowledgment}

The authors would like to thank the James Dyson Foundation for funding this research. The authors would also like to thank Dr. Simon Box of the University of Southampton, who has provided expertise on studying problems in the domain of traffic research, and advice on how to utilise the SUMO API (TraCI).

\bibliographystyle{ieeetr}
\bibliography{library}

\end{document}